\documentclass[letterpaper]{jpconf}
\begin{document}
\title{Towards an Understanding of the New Charm and Charm-Strange Mesons}

\author{Stephen Godfrey}

\address{Ottawa-Carleton Institute for Physics, \\
Department of Physics, Carleton University, Ottawa, Canada K1S 5B6}

\ead{godfrey@physics.carleton.ca}

\begin{abstract}
The observation of the 
$D_{sJ}^*(2317)$, $D_{sJ}(2460)$, and SELEX $D^*_{sJ}(2632)$
states with properties differing considerably from what was expected 
has led to a renewed interest in hadron spectroscopy.  In addition to 
these states, non-strange partners of the $D_{sJ}$ states have also 
been observed.  Understanding the $D_0^*$ and $D_1'$ states can provide 
important insights into the $D_{sJ}$ states.  In this contribution I 
examine quark model predictions for the $D_0^*$ and $D_1'$ states and 
discuss experimental measurements that can shed light on them. 
I find that these states are well described 
as the broad, $j=1/2$ non-strange charmed $P$-wave mesons.
In the latter part of this writeup I discuss the $c\bar{s}$ 
possibilities for the 
SELEX $D^*_{sJ}(2632)$ and measurements that can shed light on it.
\end{abstract}.

\section{Introduction}

The last eighteen months have been exciting times for hadron 
spectroscopists with the first observation of many charmed and 
charmonium states. It started with the observation of the 
$D_{sJ}^*(2317)$ \cite{Aubert:2003fg} which was described as having 
properties ``quite different from those predicted by quark potential 
models''.  The $D_{sJ}(2460)$\cite{Besson:2003cp} was observed shortly 
thereafter with similar discrepancies between theory and experiment.
To understand the nature of the discrepancies we note that 
the four $L=1$ $P$-wave mesons can be grouped into two doublets 
characterized by the angular momentum of the light quark; $j=3/2$, $1/2$.  
The $j=3/2$ $c\bar{s}$ states were predicted to be relatively narrow 
and are identified with the $D_{s1}(2536)$ and $D_{s2}(2573)$ states
while the $D_{s0}^*$ and $D_{s1}'$ $j=1/2$ states were expected to
have large $S$-wave widths decaying to 
$DK$ and $D^*K$ respectively\cite{Godfrey:1986wj}.  
However, the states observed by 
Babar and CLEO are below $D^{(*)}K$ threshold and are very narrow. 
This has led to considerable theoretical speculation that these states 
may be something new such as multiquark states or meson-molecules
\cite{Colangelo:2004vu}.  My 
view is that the $D_{sJ}^{(*)}$ states
are conventional $q\bar{q}$ states with their masses shifted due to 
coupling to the nearby $DK$ and $D^*K$ open channels.  
Diagnostic tests have been proposed to help
understand the nature of these newly discovered states
\cite{Colangelo:2004vu,Godfrey:2003kg,Bardeen:2003kt,Colangelo:2003vg}.  
In addition,
the non-strange $j=1/2$ $P$-wave states, $D_0^*$ and $D_1'$, 
have been observed \cite{Abe:2003zm,Link:2003bd,Anderson:1999wn}
and 
comparing their observed properties with 
theoretical predictions can give us some sense of how reliable the 
models are \cite{Colangelo:2004vu,Godfrey:2004ct}.  
The first part of this writeup examines quark model 
predictions of properties of the charm and charm-strange $P$-wave 
mesons and some diagnostic tests of the $q\bar{q}$ explanation.

The most recent addition to the family of charm meson misfits is the 
$D_{sj}^*(2632)$ state observed by the SELEX collaboration 
\cite{Evdokimov:2004iy}.  Again, there has been considerable 
speculation about what this state might be.  In the second part of 
this writeup I discuss the quark model possibilities and outline  
some measurements that can be used to test them \cite{Barnes:2004ay,Chao:2004nb}.

\section{The Charmed P-wave Mesons}

Almost all the theoretical effort has been devoted to explain the 
$D_{sJ}^{(*)}$ states.  The non-strange partners 
have received almost no attention although they also contain 
important spectroscopic information and could hold the key to 
understanding the $D_{sJ}^{(*)}$'s or at least tell us how reliable 
our models are.  

The measured properties of the $L=1$ charm mesons are summarized in 
Table 1 along with quark model 
predictions \cite{Godfrey:1986wj,Godfrey:2004ct,Godfrey:1985xj}. 
The quark model gives a 
$P$-wave cog that is $\sim 40$~MeV 
too high but the splittings are in very good
agreement with the measured masses.  The width predictions are given 
for the pseudoscalar emission model with the flux-tube 
model giving qualitatively similar results \cite{Godfrey:1986wj}.
We note that Belle \cite{Abe:2003zm} 
and FOCUS \cite{Link:2003bd} measure 
$\Gamma(D_2^{*0})=37\pm 4.0$~MeV and $\Gamma(D_1^0)=23.7\pm 4.8$~MeV 
which are larger than the PDG values.  They attribute 
differences with older results to taking into account interference with 
the broader $D$ states.  Overall the agreement between theory and 
experiment is quite good.  Note that the physical $D_1^{(\prime)}$ 
states are linear combinations of the $^3P_1$ and $^1P_1$ states so 
that the good agreement for the decay widths reflects a successful 
prediction for the $1^3P_1 -1^1P_1$ mixing angle.

\begin{table}[h]
\caption{Comparison of Quark Model Predictions$^{3,11,15}$
to Experiment for the $L$=1 Charm Mesons.}
\begin{center}
{\begin{tabular}{@{}ccccc@{}} 
\br
State & \multicolumn{2}{c}{Mass (MeV)}  & \multicolumn{2}{c}{Width (MeV)} \\
	& Theory$^a$ & Expt & Theory$^{b,7,10}$ & Expt \\
\mr
$D_2^*$ & 2460 & $2459\pm 2$ $^c$  & 54 & $23\pm 5$ $^c$\\
$D_1$   & 2418 & $2422\pm 1.8$ $^c$& 24 & $18.9^{+4.6}_{-3.5}$ $^c$ \\
$D_1'$  & 2428 & $2438\pm 30$ $^d$ & 250  & $329\pm 84$ $^d$\\
$D_0^*$ & 2357 & $2308\pm 36$ $^e$ & 280 & $276\pm 66$ $^e$\\
\br
\end{tabular}}
\end{center}
$^a$ The $P$-wave cog \cite{Godfrey:1986wj,Godfrey:1985xj} 
was shifted down 42~MeV. \\
$^b$ Using the masses from column 2. \\
$^c$ Particle Data Group \cite{Eidelman:2004wy} \\
$^d$ Average of the Belle \cite{Abe:2003zm} 
and CLEO \cite{Anderson:1999wn} $D_1'^{0}$ measurements\\
$^e$ Belle Collaboration \cite{Abe:2003zm} 
\end{table}

Radiative transitions probe the internal structure of 
hadrons \cite{Godfrey:2003kg,Bardeen:2003kt,Colangelo:2003vg}. 
Table 2 gives the quark model predictions for 
E1 radiative transitions between the $1P$ and $1S$ charm 
mesons \cite{Godfrey:2004ct}.  The $D^{*0}_2\to D^{*0} \gamma$, 
$D^{0}_1\to D^{*0} \gamma$ and $D_1^0\to D^{0}\gamma$ 
transitions should be observable.  The latter two are of particular 
interest since the ratio of these partial widths measure the 
$^3P_1-^1P_1$ mixing angle in the charm meson sector which is a good test 
of how well the Heavy Quark Limit is satisfied.

\begin{table}[h]
\caption{Partial widths and branching ratios for 
 E1 transitions between $1P$ and $1S$ charm mesons.  
The widths are given in keV unless otherwise noted.  
The $M_i$ and the total widths used to calculate the BR's 
are taken from Table 1.  The matrix elements are calculated using the 
wavefunctions of Ref. 15.
}
\begin{center}
{\begin{tabular}{@{}l l c c c c c c@{}} 
\br
Initial & Final & $M_i$ & $M_f$ &  $k$ & 
	$\langle 1P | r | nS \rangle $ &  Width  & BR  \\
state  & state & (GeV) & (GeV) & (MeV) & (GeV$^{-1}$) & (keV) & \\
\mr
$D_{2}^{*+}$ & $D^{*+} \gamma $ & 2.459 & 2.010 & 408 & 2.367 & 57 & 0.25\% \\
$D_{2}^{*0}$ & $D^{*0} \gamma $ & 2.459 & 2.007 & 411 & 2.367 & 559 & 2.4\% \\
$D_{1}^+$    & $D^{*+} \gamma$ & 2.422 & 2.010 & 377 & 2.367 & 8.8 & 
				$5\times 10^{-4}$ \\
	   & $D^{+} \gamma$ & 2.422 & 1.869 & 490 & 2.028 & 58 & 0.3\% \\
$D_{1}^0$    & $D^{*0} \gamma$ & 2.422 & 2.007 & 380 & 2.367 & 87 & 0.5\% \\
	   & $D^{0} \gamma$ & 2.422 & 1.865 & 493 & 2.028 & 571 & 3.0\% \\
$D_{1}'^+$    & $D^{*+} \gamma$ & 2.428 & 2.010 & 382 & 2.367 & 37 & 
				$10^{-4}$ \\
	   & $D^{+} \gamma$ & 2.428 & 1.869 & 494 & 2.028 & 15 & 
		$4\times 10^{-5}$ \\
$D_{1}'^0$    & $D^{*0} \gamma$ & 2.428 & 2.007 & 385 & 2.367 & 369 & 0.1\% \\
	   & $D^{0} \gamma$ & 2.428 & 1.865 & 498 & 2.028 & 144 & 
		$4\times 10^{-4}$ \\
$D_{0}^{*+}$ & $ D^{*+} \gamma$ & 2.357 & 2.010 & 321 & 2.345 & 27 & $10^{-4}$ \\
$D_{0}^{*0}$ & $ D^{*0} \gamma$ & 2.357 & 2.007 & 324 & 2.345 & 270 & 0.1\% \\
\br
\end{tabular}}
\end{center}
\end{table}

The overall conclusion is that the quark model describes the 
$P$-wave charm mesons quite well and 
models invoked to describe the $D_{sJ}^*(2317)$ and  
$D_{sJ}(2460)$ states should also explain their non-strange charm meson 
partners.  Better data would further test the models.

\section{The Charm-Strange P-wave Mesons}

Turning to the $D_{sJ}$ states, the narrow $j=3/2$ 
states are identified with the $D_{s1}(2536)$ and $D_{s2}(2573)$
with their observed properties in good 
agreement with quark model predictions\cite{Godfrey:1986wj,Godfrey:1985xj}.  
The $j=1/2$ states were predicted 
to be broad and to decay to $DK$ and $D^*K$ and were not previously 
observed.  But the $D^*_{sJ}(2317)$ is below $DK$ threshold and the 
$D_{sJ}(2460)$ is below $D^*K$ threshold so the only allowed strong 
decays are $D_{sJ}^{(*)}\to D_s^{(*)}\pi^0$ which violate isospin and 
are expected to have small widths ${\cal O}(10)$~keV 
\cite{Godfrey:2003kg,Bardeen:2003kt,Colangelo:2003vg}. 
As a consequence, the radiative 
transitions are expected to have large BR's and 
are an important diagnostic tool to understand the nature of 
these states \cite{Godfrey:2003kg,Bardeen:2003kt,Colangelo:2003vg}.  
Although there are discrepancies between the 
quark model predictions and existing measurements they can be 
accomodated by the uncertainty in theoretical estimates of 
$\Gamma(D_{sJ}^{(*)}\to D_{s}^{(*)}\pi^0)$ and 
by adjusting the $^3P_1-^1P_1$ mixing angle 
for the $D_{s1}$ states. As in the case of the $D_1$ states, the 
radiative transitions to $D_{s}$ and $D^*_s$ can be used to constrain
the $^3P_1-^1P_1$ ($c\bar{s}$) mixing angle.  

The problems with the newly found $D_{sJ}$ states are the mass 
predictions. Once the masses are fixed the narrow widths follow.  
My view is that the strong coupling to $DK$ (and $D^*K$)
is the key to solving this puzzle.  A number of people have studied 
this and have found that coupled channel effects lead to the required 
mass shifts \cite{coupled}.  
Unfortunately, coupled channel effects also appear to lead to 
comparable shifts in states that were previously in good agreement 
with experiment \cite{swanson}.  
So it is still an open question whether coupled channel 
effects can account for the discrepancy between quark model 
predictions for the $D_{sJ}$ masses and experiment.

\section{Options for the $D_{sJ}^*(2632)$} 

The SELEX collaboration recently observed a narrow charm-strange meson 
decaying into $D_s^+\eta$ and $D^0K^+$ final states in the ratio
of $\Gamma(D^0K^+)/\Gamma(D_s^+\eta) =0.16\pm 0.06$ with
$M=2632.6\pm 1.6$~MeV/c$^2$ and $\Gamma<17$~MeV/c$^2$ at 90\% C.L. 
\cite{Evdokimov:2004iy}.  A number of 
possible assignments have been suggested: a conventional $2^3S_1(c\bar{s})$ 
state, a $c\bar{s}$ hybrid, or a two-meson molecule \cite{Barnes:2004ay}.  
In this section 
we consider the conventional $c\bar{s}$ options for the $D_{sJ}^*(2632)$
\cite{Barnes:2004ay}.

The most plausible $c\bar{s}$ state is the $2^3S_1(c\bar{s})$ state 
with $M=2730$~MeV although the $1^3D_1(c\bar{s})$ is somewhat higher with a mass of 
2900~MeV \cite{Godfrey:1985xj}.  
As suggested with respect to the $D^*_{sJ}(2317)$ 
and $D_{sJ}(2460)$ states, the $q\bar{q}$ predictions 
could be shifted through mixing with the two-meson continuum. Note 
that the mass of the $K^*(1410)$ is also lower than 
quark model predictions and its partial widths also disagree 
with decay models.  So 
it is possible that the discrepancies of the $K^*(1410)$ and $D_{sJ}^*(2632)$
properties are somehow related.

The allowed open flavour decay modes of a $2^3S_1(c\bar{s})$ state with
 mass $M=2632$~MeV are $DK$, $D_s \eta$, and $D^*K$.  We found that
$\Gamma(D_{sJ}^*(2632))=36$~MeV with 
$\Gamma(D^*K)>\Gamma(DK)>>\Gamma(D_s\eta)$. In particular 
$\Gamma(DK)/\Gamma(D_s\eta)\simeq 9$.  For comparison, SELEX finds
$\Gamma(D^0K^+)/\Gamma(D_s^+\eta) =0.32\pm 0.12$, (taking into account 
$D^0K$ and $D^+K^0$).  Clearly there is an inconsistency.  
$2^3S_1-1^3D_1$ mixing, which could be generated by coupling to decay 
channels, could alter this ratio but it would require both fine tuning of 
the mixing angle and of a quark model parameter to an unlikely value.  
Although this tuning cannot be ruled out we consider it unlikely.  We 
further estimate $\Gamma(2^3S_1(c\bar{s})\to D_s^*\pi\pi)\simeq 
220$~keV implying BR $\geq 1\%$.  If the SELEX state is indeed the  
$2^3S_1(c\bar{s})$ state the $D^*K$ decay mode must be seen.
The  $D_{sJ}^*(2632)$ 
should be seen in $B$-decay, the $D_s\pi\pi$ mode should be present 
with BR $\geq 1\%$, and the $1^3D_1$ state should be $\sim$200~MeV 
higher in mass. 

\section{Summary}

To summarize, we found that the $P$-wave charm mesons are well 
described by the quark model.  However, it is important to confirm the 
broad $j=1/2$ states and obtain more precise measurements of their 
properties.  The $D^*_{sJ}(2317)$ and $D_{sJ}(2460)$ states have 
masses lower than expected for the missing $0^+$ and $1^+$ $j=1/2$ 
$c\bar{s}$ states.  This may be due to coupling to decay channels but 
further work is needed.  In any case,
radiative transitions are a good way of testing the nature of these 
states.

If the SELEX $D_{sJ}(2632)$ state is the  $2^3S_1(c\bar{s})$ state it 
should decay to $D^*K$ with a sizable branching ratio.  The 
$2^3S_1(c\bar{s}) \to D_s^* + \pi\pi$ decay mode should also be present 
with a partial width of $\sim 220$~keV and a BR of $\sim 1$\%. 
We expect that 
the  $2^3S_1(c\bar{s})$ should be seen in $B$-decays.  The $1^3D_1(c\bar{s})$
should also be present with a mass roughly 200~MeV higher.  We 
encourage experimenters to search for these states in $B$-decay.

\section*{Acknowledgments}
This work was supported by the Natural Sciences and 
Engineering Research Council of Canada.


\section*{References}

\begin{thebibliography}{9}


\bibitem{Aubert:2003fg}
B.~Aubert {\it et al.}  [BABAR Collaboration],
Phys.\ Rev.\ Lett.\  {\bf 90}, 242001 (2003).

\bibitem{Besson:2003cp}
D.~Besson {\it et al.}  [CLEO Collaboration],
Phys.\ Rev.\ D {\bf 68}, 032002 (2003).

\bibitem{Godfrey:1986wj}
S.~Godfrey and R.~Kokoski,
Phys.\ Rev.\ D {\bf 43}, 1679 (1991).

\bibitem{Colangelo:2004vu}
For a recent review see 
P.~Colangelo, F.~De Fazio and R.~Ferrandes,
Mod.\ Phys.\ Lett.\ A {\bf 19}, 2083 (2004).


\bibitem{Godfrey:2003kg}
S.~Godfrey,
Phys.\ Lett.\ B {\bf 568}, 254 (2003).

\bibitem{Bardeen:2003kt}
W.~A.~Bardeen, E.~J.~Eichten and C.~T.~Hill,
Phys.\ Rev.\ D {\bf 68}, 054024 (2003).

\bibitem{Colangelo:2003vg}
P.~Colangelo and F.~De Fazio,
Phys.\ Lett.\ B {\bf 570}, 180 (2003).

\bibitem{Abe:2003zm}
K.~Abe {\it et al.}  [Belle Collaboration],
Phys.\ Rev.\ D {\bf 69}, 112002 (2004).

\bibitem{Link:2003bd}
J.~M.~Link {\it et al.}  [FOCUS Collaboration],
Phys.\ Lett.\ B {\bf 586}, 11 (2004).

\bibitem{Anderson:1999wn}
S.~Anderson {\it et al.}  [CLEO Collaboration],
Nucl.\ Phys.\ A {\bf 663}, 647 (2000).

\bibitem{Godfrey:2004ct}
S.~Godfrey,
[hep-ph/0409236];
S.~Godfrey, in preparation.

\bibitem{Evdokimov:2004iy}
A.~V.~Evdokimov {\it et al.}  [SELEX Collaboration],
hep-ex/0406045.

\bibitem{Barnes:2004ay}
T.~Barnes, F.~E.~Close, J.~J.~Dudek, S.~Godfrey and E.~S.~Swanson,
Phys.\ Lett.\ B {\bf 600}, 223 (2004). 

\bibitem{Chao:2004nb}
K.~T.~Chao,
Phys.\ Lett.\ B {\bf 599}, 43 (2004).

\bibitem{Godfrey:1985xj}
S.~Godfrey and N.~Isgur,
Phys.\ Rev.\ D {\bf 32}, 189 (1985).

\bibitem{Eidelman:2004wy}
S.~Eidelman {\it et al.}  [Particle Data Group Collaboration],
Phys.\ Lett.\ B {\bf 592}, 1 (2004).

\bibitem{coupled}
E.~van Beveren and G.~Rupp,
Phys.\ Rev.\ Lett.\  {\bf 91}, 012003 (2003);
%
D.~S.~Hwang and D.~W.~Kim,
Phys.\ Lett.\ B {\bf 601}, 137 (2004);
%
Y.~A.~Simonov and J.~A.~Tjon,
Phys.\ Rev.\ D {\bf 70}, 114013 (2004);
%
D.~Becirevic, S.~Fajfer and S.~Prelovsek,
Phys.\ Lett.\ B {\bf 599}, 55 (2004).

\bibitem{swanson}
E. Swanson, private communication.

\end{thebibliography}
\end{document}